\documentclass[aps,prd,twocolumn, nofootinbib,floatfix]{revtex4} 
\def\diag{\mbox{diag}}

\usepackage{epsfig} 
 
\def\gsim{  \lower .75ex \hbox{$\sim$} \llap{\raise .27ex \hbox{$>$}} }  
\def\lsim{  \lower .75ex \hbox{$\sim$} \llap{\raise .27ex \hbox{$<$}} } 
\def\be{\begin{equation}} 
\def\ee{\end{equation}}

\def\N{{\cal N}} 
 
\def\erf{\mbox{erf}} 
 
\def\mpl{M_{Pl}}

\def\bphi{\bar{\phi}}

\begin{document} 
 
\title{Cosmology From Random Multifield Potentials}

\author{Amir Aazami} 
\author{Richard Easther}

\affiliation{Department of Physics, Yale University, New Haven  CT 06520, USA}

\begin{abstract} 
We consider the statistical properties of vacua and inflationary trajectories associated with a random multifield potential.  Our underlying motivation is the string landscape, but our calculations apply to general potentials. Using random matrix theory, we  analyze the Hessian matrices associated with the extrema of this potential. These potentials generically have a vast number of extrema. If the  cross-couplings (off-diagonal terms) are of the same order as  the self-couplings  (diagonal terms) we show that essentially all extrema are saddles, and the number of minima is effectively zero. Avoiding this requires the same separation of scales needed to ensure that Newton's constant is stable against radiative corrections in a string landscape.   Using  the central limit theorem we find that even if the number of extrema is enormous, the typical distance between extrema is still substantial -- with challenging implications for inflationary models that depend on the existence of a complicated path inside the landscape. 
\end{abstract} 
 
\maketitle 

\section{Introduction}

We study the statistical properties of potentials with many  degrees of freedom, and the inflationary trajectories that exist inside them. A large part of our interest in this problem follows from the  string landscape \cite{Bousso:2000xa,Kachru:2003aw,Susskind:2003kw}.  However, we focus on the statistical properties of arbitrary multifield potentials, without making direct appeals to stringy physics or supersymmetry.     Consequently, there is no guarantee that the string landscape will conform to the conclusions we reach. Rather, this paper determines  what we can expect {\em solely\/} on  the basis of the string landscape being  a large, multidimensional potential.\footnote{We use the term ``landscape'' to refer to any potential with  large numbers of scalar degrees of freedom -- whereas the phrase ``string landscape'' denotes the {\em actual\/} landscape (if any) that is provided to us by string theory.}  In doing so we have two aims. Firstly, to the extent that the properties of the string landscape are a function solely of its dimensionality and other simple parameters, the analysis presented here elucidates  its properties.  Secondly, if the actual string landscape does not match the conclusions we reach here, we will learn that its properties  follow from the structure of string theory itself, and not just large-$N$ statistics.

A program for computing the number of vacua in the string landscape has been implemented \cite{Douglas:2003um,Ashok:2003gk,Denef:2004ze,Giryavets:2004zr}. However,  in keeping with our interest in what follows from the large dimensionality of the landscape, our starting point will be an elegant argument championed by Susskind \cite{Susskindtalk}, which explains why we expect a landscape  to possess a huge number of extrema, and a smaller but still vast number of minima.  We focus on three specific problems -- the number of minima (relative to the number of extrema) that survive when cross-couplings between the individual fields are turned on,  the average separation between extrema in a random landscape, and the inflationary trajectories that exist within the landscape.  Calculations with a  similar flavor to those presented here can be found in Arkani-Hamed {\it et al.\/}'s work on the ``friendly landscape''  \cite{Arkani-Hamed:2005yv}, extended by Distler and  Varadarajan to cross-coupled fields through the use of random polynomials and modular forms \cite{Distler:2005hi}.    The latter paper, in particular, examines a very similar problem to ours, using algebro-differential techniques to analyze the minima of the landscape.    Here, we focus upon the random matrices arising from the Hessians  which characterize the extrema of a random cross-coupled potential.  We  find that the cross-coupling terms must be suppressed relative to the self-interaction terms by  $\sim 1/\sqrt{N}$, where $N$ is the dimensionality of the landscape, or the number of minima is effectively reduced to zero.   This coincides (up to a numerical factor) with the separation of scales needed to ensure that Newton's  constant is safe from radiative corrections  \cite{Arkani-Hamed:2005yv}.  We also look at higher order corrections, showing that these can induce an additional $\log{(N)}$ term in this result.      Further, random matrix theory models of the landscape are considered by \cite{Kobakhidze:2004gm,Mersini-Houghton:2005im,Holman:2005ei}, in the context of quantum cosmology.

In addition to estimating the number of vacua permitted by a random multifield potential, we also consider the very large number of inflationary trajectories  associated with a random landscape. Slow roll inflation will typically occur near a saddle, rather than at a minimum -- and  saddles are much more numerous than minima or maxima.  Many inflationary models have been built inside of string theory (e.g. \cite{Banks:1995dp,Binetruy:1996xj,Dvali:1998pa,Burgess:2001fx,Burgess:2001vr,Jones:2002cv,Arkani-Hamed:2003mz,Kachru:2003sx,Silverstein:2003hf,Alishahiha:2004eh,Hsu:2003cy,Firouzjahi:2003zy,Hsu:2004hi,Iizuka:2004ct,DeWolfe:2004qx,Dvali:2003vv,Easther:2004qs,Blanco-Pillado:2004ns,Easther:2004ir,Freese:2004vs,Burgess:2005sb,McAllister:2005mq,Cremades:2005ir}), and a number of these make explicit reference to the properties of the landscape.  In particular, folded inflation \cite{Easther:2004ir} and chained inflation \cite{Freese:2004vs}, and  rely on the presumed combinatorics of the landscape  to concatenate several successive periods of inflation and we will see that a {\em generic\/} landscape is not likely to have the features these models need. Random models of inflation have been considered in other contexts.   Monte Carlo Reconstruction provides  a mechanism for generating an arbitrarily large number of random inflationary models, via the Hubble Slow Roll formalism \cite{Easther:2002rw}. Moreover, given the extension of the Hubble Slow Roll formalism to multifield models, the generalization of Monte Carlo reconstruction to more general potentials would be straightforward \cite{Easther:2005nh} although of limited value given the large number of free parameters this necessarily involves.   In separate work, Tegmark has looked closely at the distribution of observable parameters associated with  random inflationary models \cite{Tegmark:2004qd}.        Finally, the problem tackled here has obvious analogues in condensed matter system -- see, for example, \cite{Fyo}.

In Section 2, we  review the expected number of extrema in a random landscape, and the proportion of these extrema which are actually minima.  We then consider the impact of cross-couplings on the distribution of minima in two limits. When the cross-couplings are of the same order as the self-couplings, we use random matrix theory to show that the number of minima is essentially zero -- or one, if we have {\em a priori\/} knowledge that the potential is bounded below. Conversely, for small cross-couplings we show that if the typical mass-scale in the landscape, $M$, satisfies $M \lesssim \mpl/\sqrt{N}$ the minima are safe. In Section 3 we review the inflationary dynamics and perturbation spectra of multifield models, and their implications for the landscape. We show that if these potentials are bounded in field-space, a weak version of the $\eta$ problem appears even without reference to supersymmetry breaking, and that viable models of inflation in a landscape are  likely to involve either an inflaton rolling toward a saddle (hybrid inflation) or several scalar fields working cooperatively to drive an inflationary epoch (assisted inflation).  In Section 4 we compute the average distance between extrema in the landscape and show that this is typically larger than $\mpl$, showing even though the landscape can contain a vast number of minima, they are typically well separated.

\section{Extrema of Random Landscapes}
\subsection{Counting and Classifying Extrema}

Susskind has given a simple geometrical argument that the number of minima in the string landscape is most likely very large  \cite{Susskindtalk}. Consider a landscape with $N$ scalar degrees of freedom, $\phi_i$, $i=1,\cdots,N$, and the functional form
\begin{equation}\label{landscape1}
L(\phi) = \sum_i f_i(\phi_i) \, ,
\end{equation}
where the $f_i$ each have at least one extremum.  Since there are no cross terms, any combination of $\phi_i$ that extremizes all the $f_i$  yields an extremum of $L$. If each $f_i$ has $\alpha_i$ extrema,  the total number of extrema of $L$ is $\prod_i \alpha_i =  \alpha^N$, where $\alpha$ is the geometric mean of the $\alpha_i$.   Continuity requires that every second extremum of $f_i$ is a minimum, so the landscape has $(\alpha/2)^N$ vacua.    In the string landscape, $N$ is expected to be on the order of a few hundred. Provided $\alpha$ is not too close to unity,  $\alpha^N \gg 10^{120}$, tempting one to use  anthropic arguments to ``explain'' why the cosmological constant, $\Lambda$, is apparently fine-tuned to this level of precision. Moreover,  when $\alpha^N \gg 10^{120}$,  the possible values of $\Lambda$ are effectively continuous (assuming they are not massively degenerate), yielding a {\em discretuum\/} of values for $\Lambda$.\footnote{Although even if $\Lambda$ does scan through some range of values, there is no guarantee that this range includes zero \cite{Arkani-Hamed:2005yv}.}  In this case minima are so closely spaced that no conceivable measurement of $\Lambda$ alone could uniquely determine which  vacuum corresponded to our universe, even if we were able to exhaustively catalog all the minima in the landscape.

Recalling elementary multivariable calculus,  $V$  possesses an extremum if all first derivatives vanish at a point.  Extrema  are classified via the Hessian, 
\begin{equation}
H = \left( \begin{array}{ccc} 
 \frac{\partial^2 V}{\partial \phi_1^2} & \cdots &  \frac{\partial^2 V}{\partial \phi_1 \partial \phi_N} \\
 \vdots & & \vdots \\
 \frac{\partial^2 V}{\partial \phi_1 \partial \phi_N} & \cdots &  \frac{\partial^2 V}{\partial \phi_N^2} 
\end{array}
\right) \, .
\end{equation}
If $H$ only has positive (negative) eigenvalues, we have a local minimum (maximum), whereas if $H$ has mixed positive and negative eigenvalues, we have a saddle point.   

\subsection{Extrema and Cross-coupled Fields}

\begin{figure} 
\includegraphics[width=3in]{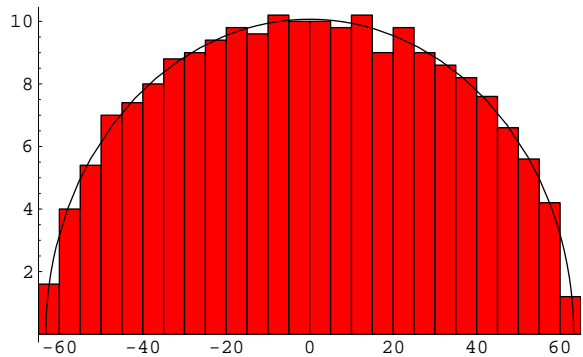}
\caption{\label{fig:histo}
The eigenvalue distribution of a $1000 \times 1000$ symmetric matrix with entries drawn from a standard normal distribution is shown, with the theoretical distribution plotted for reference.   }
\end{figure}

\begin{figure}
\includegraphics[width=3in]{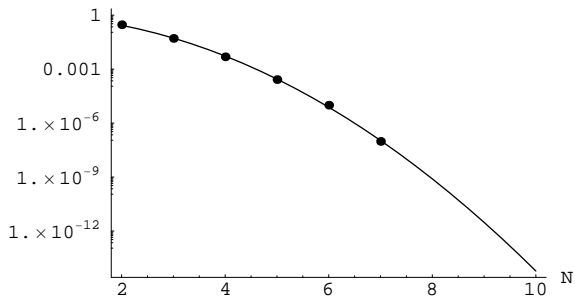}
\caption{\label{fig:fit}We show the proportion, $f$ of randomly generated symmetric $N \times N$ matrices for which all eigenvalues have the same sign.  The curve is well fit by $\sim \exp(- b N^2)$, with  the key observation being that $f$ is decreasing faster than exponentially with $N$. }
\end{figure}

Since $V$ is a function with random coefficients, $H$ is a random matrix.   The theory of random matrices is rich and beautiful, with applications all over physics \cite{Mehta}.   The simplest case, where the landscape is defined by equation~(\ref{landscape1}), is trivial: $H$ is diagonal, and the eigenvalues are the second derivatives of the $f_i$ at a given extremum. These are equally likely to be negative  or positive, so we have $2^N$ equally  weighted permutations of the signs of the eigenvalues. Only one of these configurations  corresponds to a minimum, and almost any randomly selected extremum will be a saddle.   Thus, of the $\alpha^N$ extrema,  $\alpha^N / 2^N = (\alpha/2)^N$ are minima, recovering the result of the previous subsection.

From a  physical perspective, however, there is no reason to expect that  the cross-terms in the landscape vanish, so let us add simple cross-couplings. Near an extremum, these take the form 
\begin{equation}\label{landscape2}
L(\phi) = \sum_i f_i(\phi_i) + \sum_{i \ne j} \epsilon_{ij} \phi_i \phi_j 
\end{equation}
We can now consider the opposite, and equally unphysical limit, where all the terms in $H$ are of equal size, and drawn from identical, independent distributions.  In this case, we need to turn to a more sophisticated analysis.  If  the $\epsilon_{ij}$  are drawn from a  normal distribution, our Hessian is drawn from the Gaussian Orthogonal Ensemble, and the resulting eigenvalues obey the Wigner semi-circle law  \cite{Mehta}.    The density of eigenvalues,\footnote{These matrices are constructed by taking a random matrix with elements drawn from a Gaussian of unit width  and then symmetrizing.  Changing the width of the distribution rescales $x$.} $E(x)$,  is
\begin{equation}\label{wigner}
E(x) = \left\{
\begin{array}{ll}
\frac{1}{\pi } \left[2 N - x^2\right]^{1/2}  \, ,&  |x|< \sqrt{2 N} \\
0 \, , & |x|> \sqrt{2 N}
\end{array}
\right.
\end{equation}
where the normalization ensures that when $E(\lambda)$ is integrated over all values of $\lambda$ the total number of eigenvalues is  $N$.     Figure~\ref{fig:histo} shows the eigenvalue distribution for a specific $1000^2$ matrix.  For the Gaussian Orthogonal Ensemble, the joint probability distribution for the eigenvalues is
\begin{equation} \label{measure}
P_N(x_1,\cdots,x_N) = C_N \exp(-{1 \over 2} \sum x_i^2 ) \prod_{j<k} |x_j - x_k|
\end{equation}
where $C_N$ is a normalization factor and $P$ gives the likelihood of finding drawing a random matrix with the eigenvalues $\{x_1,\cdots,x_N\}$  \cite{Mehta}.  For our purposes, the key feature of this distribution is that it is ``rigid'' -- fluctuations in the eigenvalue distribution are very small.  Consequently, the likelihood of a large fluctuation putting all the eigenvalues on the same side of zero is vastly smaller than $1/2^N$.  

A great deal is known about the properties of fluctuations away from the semi-circle law (e.g. \cite{Tracy:1995xi}) and we can  estimate the likelihood of all the eigenvalues having the same sign.  Firstly, change variables from $x_i$ to $a_i = x_i - x_{i-1}, a_1 = x_1$. Without loss of generality we can assume that   $x_1<\cdots< x_N$, so the $a_i$ are all positive for $i>1$ and we can drop  the absolute value symbol in equation~(\ref{measure}). The probability that all the eigenvalues are positive is
\begin{equation} \label{weight1}
 p(x_i >0 , \forall i) = \frac{\int_0^\infty da_1 \int_0^\infty da_2 \cdots \int_0^\infty  da_N 
     P_N(a_i)}{\int_{-\infty}^\infty da_1 \int_0^\infty da_2 \cdots \int_0^\infty  da_N  P_N(a_i)}  \, .
 \end{equation}
The integrals over $a_i$, $i\ge2$ necessarily have lower bounds of zero -- whereas all the eigenvalues must be positive if $a_1>0$, and  the integral over $a_1$ thus starts at $-\infty$ in the denominator, and $0$ in the numerator. Performing {\em only\/} the integrals over $a_1$, and leaving the positions of the other eigenvalues unspecified, after some algebra and a great deal of cancellation we find
\begin{equation} \label{weight2}
  p(a_2,\cdots,a_N | a_1 \ge 0)  = \frac{1}{2} \left[1 - \erf\left( \sum_{i=1}^{N-1} \frac{a_{i+1} (N-i)}{\sqrt{2N}} \right) \right] \, .
\end{equation}
This calculation is exact up to this point. We now guess that the average $a_i$ is given by assuming that the eigenvalues are uniformly distributed between $(0,\sqrt{2N})$ so $a_i \sim \sqrt{N/2}$.  After making this brutal  approximation\footnote{This is crude for two reasons. Firstly, the $a_i$ are not all the same, and secondly the maximal eigenvalue is not likely to be exactly $\sqrt{2N}$ after a fluctuation this large.}  and employing the asymptotic form of the error function,  we find
 \begin{equation} \label{weight3}
  p(x_i >0 , \forall i)  \sim \frac{1}{2} \exp{\left(-\frac{N^2}{4}\right)} \frac{\sqrt{\pi}}{N}
  \end{equation}

In Figure~(\ref{fig:fit}), we have plotted $p(\lambda_i >0 , \forall i)$ for randomly generated symmetric $N\times  N$ matrices. This data is accurately fitted by  the heuristic relationship
\begin{equation}
\log{p} = 
 - 0.32491 N^{-2.00387} \, .
\end{equation}
Despite the approximation used to obtain equation~(\ref{weight3}),  we have confirmed that the likelihood that all the eigenvalues  of an $N \times N$ symmetric matrix have the same sign scales as $e^{-c N^2}$. The measured constant differs slightly from $-0.25$, although given the simplicity of our approximation the agreement is perhaps surprisingly good.  

Tracy and Widom have proved very general results about the expected size of the largest eigenvalue, and the likelihood of fluctuations away from the expected maximal value \cite{Tracy:1995xi}.   The likelihood of finding that the largest eigenvalue is zero is of course equivalent to the probability that all the eigenvalues have the same sign.  In principal, one could extract an exact formula for this ratio at large $N$ from \cite{Tracy:1995xi}, but since we are working with a very simple model, a more sophisticated calculation is probably not justified.

The strong suppression of large fluctuations away from the semi-circle law can be understood physically. The eigenvalues of a symmetric matrix are physically equivalent to the distribution of charged beads along an infinite wire with a quadratic potential centered on the origin, and logarithmic interaction potentials between the beads.   The overall potential causes the beads to cluster near the origin, but their mutual repulsion forces them to separate \cite{Mehta}. The likelihood of  a fluctuation that places all the beads on the same side of the origin is very small, and  vastly less than the $1/2^N$ likelihood that the eigenvalues of a diagonal matrix will all have  the same sign.  

The number of extrema in the landscape  was calculated after assuming that the fields are uncoupled, so that the number of extrema encountered as one adjusts a single field does not depend on the value of the other fields.  We are now tacitly assuming that even in a cross-coupled landscape the number of extrema depends only exponentially on $N$.  The precise form of this dependence does not matter, since the chance of any given extremum being a minimum (or maximum) now decays super-exponentially with increasing $N$.  In this limit, instead of  having $\gg 10^{120}$ vacua, the potential  has none -- or perhaps just one, if it is bounded below.  In a landscape with cross-couplings of the same order as the diagonal terms, the field point will never be prevented from rolling and will naturally flow to the global minimum (assuming that one exists).  While such a potential yields a universe that is now in a unique and stable vacuum, one cannot use combinatorial arguments to dilute the cosmological constant problem.

\subsection{Landscapes With Small  Cross-couplings}

In reality, it is reasonable to assume that the landscape can be decomposed into two pieces  -- a diagonal part corresponding to the $f_i(\phi)$ which are of order unity, and much smaller cross-terms of order $\epsilon$. Physically, the diagonal terms are self-interactions and the cross terms come from interactions between the individual fields.  If  the characteristic mass-scale of the tree level $f_i(\phi_i)$   is $M$ then $\epsilon$ has dimensions of mass-squared, and we expect  $\epsilon \sim M^4 /\mpl^2$, in the absence of special symmetries and assuming that the cross-terms are induced by gravitational interactions.

If $\epsilon \sim 1$ (that  is, the cross-couplings arise from operators of the same order as those that produce the diagonal terms),  we have established that almost any extremum will be a saddle.  We now determine how large the cross coupling terms can be  without destroying the expectation that  $2^{-N}$ of the extrema are actually minima.  

We write the Hessian as $H = X+ U$, where $X = \diag(x_1,x_2,\dots,x_N)$ is a diagonal matrix with distinct, positive eigenvalues $x_l$ and  $U$ is an $N \times N$ symmetric matrix.\footnote{If the eigenvalues of $X$ are not distinct the problem becomes a little more complicated, but will not affect out conclusions here.}  The eigenvalues of $H$, $\lambda_l$ can be written down as a perturbation series using Moore-Penrose inverses \cite{Schott}, and up to third-order in the elements of $U$, 
\begin{eqnarray} 
\lambda_l &\approx& x_l + u_{ll} - \sum_{i \neq l} {u_{il}^2 \over (x_i - x_l)} - 
 \sum_{i \neq l} {u_{ll}u_{il}^2 \over (x_i - x_l)^2} +  \nonumber \\  
 && \sum_{i \neq l} \sum_{j \neq l} {{u_{il}u_{jl}u_{ij} \over (x_i - x_l)(x_j - x_l)}} \, . \label{perturb1}
\end{eqnarray}
Assume our Hessian is associated with a minimum of the diagonal landscape $X$.    If the average eigenvalue of $X$ is on the order of $1$, then we guess that the lowest eigenvalue of $X$ is roughly $1/N$ at a local minimum. We immediately see that only the lowest lying eigenvalues of $X$ are at risk of having their signs  changed by adding the cross-coupling terms.  

Near a minimum, the $\lambda_i$ are  the masses of the $N$ scalar degrees of freedom describing oscillations away from that minimum.   At lowest order, these are given by the $x_l$.  The  diagonal terms of the symmetric contribution, $U$, correspond to the loop corrections to these masses.  To prevent the mass corrections from changing the sign of the lowest eigenvalue, the standard deviation $u_{ll}$ must be (at least) several times smaller than $1/N$.\footnote{We assume that the entries of $U$ and $X$ are uncorrelated, and that all the $u_{ij}$ are drawn from similar distributions.}  For definiteness, we express this bound as $|u| \sim \epsilon\lesssim 1/(3 N)$ -- where the $3$ is chosen so that a 3$\sigma$ term in $U$ has the ability to change the sign of an eigenvalue.\footnote{The ``3" here should really be obtained from an expansion of the tail of the integrated Gaussian distribution. However, since a 4$\sigma$ fluctuation is 17 times rarer than a 3$\sigma$ fluctuation (and the ratio increases as we move further from the mean), this is a very mild correction and we ignore it.}    

If we assume the entries of $U$  are induced by Planck scale operators, we can relate the dimensionality of the landscape to a separation of scales. Restoring units to $X$, $X \sim M^2$,   
\begin{eqnarray}
&&\epsilon \sim \frac{M^4}{\mpl^2} \sim \frac{M^2}{3 N} \nonumber \\ 
&\Rightarrow&  \frac{M}{\mpl} \lesssim \frac{1}{\sqrt{3N} } 
\end{eqnarray}
Interestingly, Arkani-Hamed {\em et al.\/} derived essentially the same relationship between the Planck scale and the scale of the landscape by requiring that Newton's constant be protected from radiative corrections:  they find an extra numerical factor arising from the constant term in the Einstein-Hilbert  action, which we have effectively absorbed into $\mpl$.

At this point, however, we have not examined the convergence of the series in equation~(\ref{perturb1}).  Looking at the second term, the numerator scales like $\epsilon^2$.  Assuming that the eigenvalues are evenly spaced, so that $x_i \sim i/N$, we estimate the magnitude of the sum as follows:
\begin{eqnarray}
  S &=&  - \sum_{i \neq l} {u_{il}^2 \over (x_i - x_l)} \, , \\
\Rightarrow S &\approx& -\epsilon^2 \sum_{i =2}^N {1 \over {\frac{i}{N} - \frac{1}{N}}}  \approx -N \epsilon^2\sum_{i=1}^N \frac{1}{i} \, , \\
\Rightarrow S &\sim&   -\epsilon^2 N \log{N} \, .
\end{eqnarray}
In order to the protect the lowest-lying eigenvalue, $\sim 1/N$ we need
\begin{equation}
\epsilon^2 < \frac{1}{N^2 \log{N}}  \, .
\end{equation}
This term {\em always\/} has a negative coefficient, and tends to render the lowest-lying mass term  at a given minimum negative. We now have a stronger constraint on the terms of $U$ than the $\epsilon \sim1/N$  derived above. The off-diagonal terms of $U$ now contribute, and in addition to correcting the masses they also induce a small rotation in the orthogonal modes, which are given by the eigenvectors of $X + U$, not $X$ alone.  This result was obtained using an explicit ansatz for the eigenvalue spacing of $X$.  Given a full understanding of this spacing, which would be possible in a detailed physical model, it would then be worth analyzing these logarithmic corrections carefully.  However, given the generic potentials we are studying here we will not pursue this issue further, beyond noting that the bounds on $\epsilon$ can easily be tighter than $1/N^2$.

\begin{figure}
\includegraphics[width=3in]{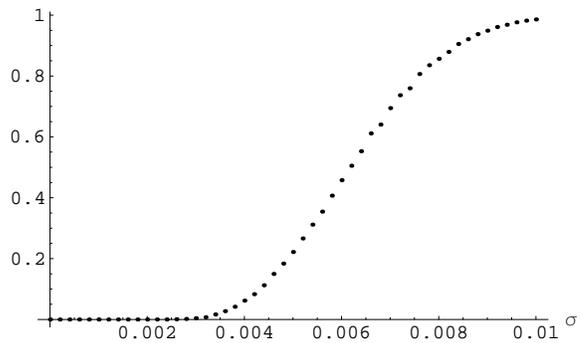}
\caption{\label{fig:perturb}     We plot the proportion of matrices $H= X+U$ whose lowest eigenvalue is  negative, where $X$ is a $\mbox{diag}(1,\cdots,N)/N$ and $U$ is a symmetric perturbation whose terms have standard deviation $\sigma$.  These results are drawn from a sample of 10000 matrices at each value of $\sigma$, with $N=100$.  }
\end{figure}

The third term of (\ref{perturb1}) is benign, since the summation  resembles $N^2 \zeta(2)$, but the extra factor of $N$ is deleted by the $u_{ll}$ term in the numerator, and we do not see a logarithmic divergence.    The final term is again safe, since its numerator takes on random signs and its contributions add incoherently.   Moreover, with this assumption we see that the above series is at least asymptotically convergent.   If   typical entries in $U$  have magnitude $1/c N$, where $c \lesssim {\cal O}(.1)$, then this series can be safely  truncated.  In Figure~\ref{fig:perturb} we demonstrate these results heuristically, showing how preserving the sign of the lowest eigenvalue depends on our choice of $\epsilon$.  

We conclude that even a mild separation of scales ensures that the off-diagonal terms in the Hessian matrix are small enough to preserve the intuition that a landscape indeed has multiple minima.   We also see that the separation of scales needed for the protection of Newton's constant \cite{Arkani-Hamed:2005yv} picks up an additional logarithmic term via this calculation.  For $N\sim1000$ this is still less than a full order  of magnitude, but it is tempting to speculate about the connection between this heuristic argument and the separation between the GUT scale and the Planck scale.   The calculation here implies an upper bound, but cannot be taken to imply that this bound is saturated in practice.  In particular, there are well-established examples (e.g. shift symmetries for axions and other Goldstone bosons, and locality in the case of fields associated with cycles that are well separated in the compact directions) where the $\epsilon$ terms are very strongly suppressed. 

The next level of sophistication, which we will not pursue here, would be to analyze a landscape where most of the fields were only weakly coupled to each other, but with each field  coupled to several others at lowest order. This yields a band-diagonal Hessian matrix. As the width of the band increased one would move from the $1/2^N$ diagonal limit to the $\exp(-N^2)$ limit found for the fully symmetric case. Given an estimate of the width of this band one could then deploy more sophisticated random matrix techniques to derive the eigenvalue distribution for the Hessian.

\section{Inflationary Dynamics}

Many discussions of the landscape focus upon the number of minima, since these are the points that are associated with the discrete vacua. Given a vast number of minima, we can at least assume the {\em existence\/} of vacua where the height of the landscape will match the observed value of cosmological constant, along with any other fundamental parameters that are effectively ``environmental.''  
There is no guarantee that these special minima are attractors, but provided they have a non-zero basin of attraction in the initial conditions space, we ``explain'' the observed value of $\Lambda$ with a very mild amount of anthropic reasoning.     From a cosmological perspective, however,  we  care about both the vacuum itself and how the universe arrives there. That is, what is the trajectory through the landscape that leads to this final (meta)-stable vacuum from a presumably random initial configuration?   In particular,  does  this trajectory include a region where the height of the landscape changes slowly, and we see a sufficiently long period of inflation?  In this case,  the combinatorial arguments advanced to ``solve'' the cosmological  constant problem may also address the tunings present in most models of inflation.

Inflation in a landscape (or with any set of scalar fields with canonical kinetic terms, for that matter) can occur in two different ways:
\begin{itemize}
\item The field point is near a minimum with $V>0$, and the fields do not possess sufficient kinetic energy to climb over the barrier.
\item The field point evolves slowly enough to ensure that $p<-\rho/3$.
\end{itemize}
The first possibility is realized today if the dark energy is generated by an actual cosmological constant, rather than some quintessence-style scenario where $w\ne-1$ or $\dot{w}\ne0$.  The second possibility is the prototype for primordial inflation, since if the field point is trapped in a local minimum it can only escape by tunneling, resurrecting the bubble problem  first recognized in the original model of inflation.  

If the Hessian is diagonal, 1 in $2^{N}$  extrema will be a minimum, but it also follows that there are $N$ in   $2^{N}$   saddles  with a {\em single} downhill direction. For successful inflation, we need the downhill direction to be relatively flat.   Near a saddle, the first derivatives necessarily vanish, and our criteria  for flatness will focus on the second derivatives in the downhill directions.

We will restrict our attention to a flat FRW universe  -- which is reasonable after {\em some\/} inflation has occurred, but avoids questions surrounding the onset of inflation.   At least locally in the landscape, we can transform the fields to ensure that they have canonical kinetic terms, so 
\begin{eqnarray}
\label{Hsqrd}
H^2&=&\frac{8 \pi G}{3}\left[{V({\phi})+\sum_{i=1}^N\frac{1}{2}\dot{\phi_i}^2}\right]  \, , \\
\frac{\ddot{a}}{a}&=&\frac{8 \pi G}{3}\left[{V({\phi})-\sum_{i=1}^N\dot{\phi_i}^2}\right]  \, ,\\
\ddot{\phi_i}&+& 3H \dot{\phi_i}+\frac{\partial V(\phi)}{ \partial \phi_i}=0 \, .
\end{eqnarray}
Here $\phi$ is a shorthand for a vector in fieldspace, $\phi = (\phi_1 ,\cdots, \phi_N)$ and labels $i$ run from $1$ to $N$, and $H$ is the usual Hubble parameter.

To sustain inflation, the kinetic term in eq.~(\ref{Hsqrd}) must  be much less than the potential term, or 
\begin{equation} 
\sum_{i=1}^N \dot{\phi_i}^2 \ll V(\phi) \, .
\end{equation}
We now stipulate
\begin{eqnarray}
 && \sum_{i=1}^N \dot{\phi_i}^2 \sim \frac{1}{8\pi G} \sum_{i=1}^N \left( \frac{V_{,i}^2}{V}\right) \ll V \\
 &\Rightarrow& \frac{1}{8\pi G} \sum_{i=1}^N \left( \frac{V_{,i}}{V}\right)^2  \ll  1
\end{eqnarray}
by requiring that $\ddot{\phi}$  is small enough to be dropped from the equations of motion.  Inflation thus occurs near points where all the first derivatives are very small or,  equivalently, in the vicinity of an extremum.\footnote{Strictly, if the first derivatives are all very small near a point, this does not {\em guarantee\/}  there is an extremum nearby. However, this qualificiation does not change the essence of our argument.}  Consequently, by counting the extrema (as opposed to just the minima)  in the landscape, we can estimate the number of distinct inflationary trajectories that can be supported within it.

In order to produce a phenomenologically acceptable period of inflation, the second derivatives of the potential must be small, at least in the downhill directions. This  is  ensures that slow roll applies not just at a point in fieldspace, but over some finite range, so that the inflationary epoch is not transient. Moreover, the first and second derivatives of the fields combine to determine the spectral index, $n_R$ of the perturbations -- but near a saddle, the spectrum is likely to be dominated by the second derivative terms.

Given our parameter $\alpha$ and the requirement that the $|\phi_i| < \mpl$ if the landscape is to be protected from large non-perturbative modifications we can  guess  the average value of $\eta = 8\pi/\mpl^2 V''/V$, assuming a single evolving field.  For definiteness, take  $V  \sim \Lambda^4 \sin( 2 \alpha \phi / \pi \mpl)$, 
\begin{equation}
\sqrt{\langle V''^2 \rangle }  = \frac{\int_{-\mpl}^{\mpl}  V''^2 }{2 \mpl}  = \frac{\Lambda^4 \pi^2 \alpha^2}{4 \sqrt{2} \mpl^2} \, .
\end{equation}
Thus the expected value of $\eta$ will be
\begin{equation}
\eta \sim  \frac{ \pi \alpha^2 }{32\sqrt{2} }  \gsim \frac{\pi}{2 \sqrt{2}} \sim 1  \, .
\end{equation}
We have taken $\alpha \gsim 4$ in the last step,  consistent with the requirement that $(\alpha/2)^N$ is a very large number.   This is a very crude version of the $\eta$ problem which plagues many supersymmetric models of inflation, although the only physical input is the requirement that the field values remain sub-Planckian.  At this level, the   $\eta$ problem does not appear to be especially onerous. A reasonable perturbation spectrum  needs $\eta \sim 0.01$, and this mild tuning could easily be supplied by relying on the vast number of potential inflationary paths.\footnote{One of us [RE] will examine the likely values of $\eta$ in single field inflation in a forthcoming paper \cite{forth1}.}  In realistic models of inflation,   $\eta$ is derived from calculable supersymmetric corrections to the potential, and these must be analysed more carefully. 

There is a hidden assumption in the analysis above: that  we are only moving in  a single direction.    However, if a landscape extends $\sim \mpl$ in any given direction, Pythagoras' theorem tells us that it measures $\sim \sqrt{N} \mpl$ from corner to corner. Consequently, if the structure of the landscape permits the field point to roll diagonally, it may traverse a distance far greater than $\mpl$. This is precisely the scenario envisaged by assisted inflation, where multiple fields act coherently to emulate a single minimally coupled scalar field \cite{Liddle:1998jc}.   This effectively embeds inflationary models with a single scalar field and  trans-Planckian vev within the landscape.     However, in order to see this, we must transform to a  new basis in which the  only downhill degree of freedom runs from one corner of the landscape to another.   If we view the landscape as a purely random potential, assisted inflation does not appear natural (the position taken in \cite{Easther:2004ir}), since it requires a correlation between the masses of a large fraction of the fields, and this reduces the effective dimensionality of the landscape, undermining the strength of combinatorial arguments. In particular   N-flation  \cite{Dimopoulos:2005ac} proposes a mechanism for realizing this type of potential within the landscape.

\section{Distance Between Extrema in the Landscape}

We now ask whether the set of extrema is large enough so that any random point will be ``near'' a minimum or saddle, and the minima and extrema thus form a dense set inside the landscape.  In the converse case, extrema -- while numerous -- still correspond to very special points inside the landscape.   We  make considerable progress by appealing to the Central Limit Theorem (see e.g. \cite{Kallenberg}). We assume the $\phi_i$ at each extremum are drawn independently from identical, flat distributions but we could relax this assumption without undermining our results. 

Take an arbitrary reference point in the landscape, $\bphi = (\bphi_1,\cdots,\bphi_N)$, and consider a set of $M$ randomly distributed points in the landscape $\phi^j = (\phi_1^j,\cdots,\phi_N^j)$ where $j\ \in  \{1,\cdots,M\}$.  In practice these points will be the set of minima or inflationary saddles.  Assume a simple Euclidean metric on the landscape, so that the distance between $\bphi$ and an arbitrary $\phi^j$ obeys
\begin{equation} \label{r2}
r_j^2  = \sum_{i=1}^N (\phi_i^j - \bphi_i)^2  \, .
\end{equation}
The Central Limit Theorem tells us that $X$,  a sum over $N$ random variables $x_i$ each with mean $\mu_i$ and variance $\sigma_i$ tends to a Gaussian as $N$ becomes large.  Constructing $X$ as follows
\begin{equation}
X  = \frac{ \sum_{i=1}^N x_i - \sum_{i=1}^N \mu_i}{\sqrt{\sum_{i=1}^N \sigma_i^2}},
\end{equation}
gives  a Gaussian with unit variance and zero mean.   Consequently, if the $\phi_i^j$ are  randomly distributed, the distribution of $r_j^2$ approaches  a Gaussian random variable.

We draw the $\phi_j$ from a uniform distribution of width $\Delta$.\footnote{Only the mean and variance of the distribution will enter the calculation, so choosing a different distribution will not change the results here dramatically.}  To keep the field values bounded by $\mpl$, we need $\Delta \lesssim \mpl$.   The  mean and variance of each term in the sum in equation~(\ref{r2}) is, 
\begin{eqnarray} \label{mu1}
\mu_i &=&\frac{1}{2\Delta} \int_{-\Delta}^\Delta d\phi_{i}^j ( \phi_i^j - \bphi_i)^2 = \frac{\Delta^2}{3} +  \bphi_i^2 \\
\sigma_i^2 &=&  \frac{1}{2\Delta} \int_{-\Delta}^\Delta  d\phi_{i}^j  [( \phi_i^j - \bphi_i)^2  - \mu_i]^2 \nonumber \\
\label{sigma1}
 &=& \frac{4}{3}  \bphi_i^2 \Delta^2 + \frac{4}{45} \Delta^4 \,.
\end{eqnarray}

Putting this together, 
\begin{equation} \label{flatR2}
R^2  =  \frac{ \sum_{i=1}^N   (\phi_i^j - \bphi_i)^2 - \sum_{i=1}^N  \frac{\Delta^2}{3} +  \bphi_i^2}{\sqrt{\sum_{i=1}^N  \frac{4}{3}  \bphi_i^2 \Delta^2 + \frac{4}{45} \Delta^4}}
\end{equation}
where $ R^2$ is a normalized version of the distribution of $r^2$.     

The above expression applies to a specified reference point in the landscape, $\bphi$. If we also choose $\bphi$ at random from the  same distribution as the $\phi^j$ we have the distribution of distances between any two randomly chosen points.   The  means and variances in  equations~(\ref{mu1}) and (\ref{sigma1}) become two dimensional integrals over both $\phi_i^j$ and $\bphi_i$, yielding
\begin{eqnarray}
\langle r^2 \rangle &=&  \frac{2}{3} N  \Delta^2 \\
\sigma(r^2)  &=& \Delta^2\sqrt{  \frac{28}{45}  N  }  
\end{eqnarray}
In the large $N$  limit, the distribution for the square of the distance between two randomly chosen points is
\begin{equation} \label{PRCLT}
P(r^2) = \frac{1}{\Delta^2}\sqrt{\frac{45}{56 \pi N} } \exp{\left[ -\frac{45(r^2 -  2\frac{N}{3}\Delta^2 )^2 }{56 N \Delta^4} \right]}.
\end{equation}
In terms of $\sigma^2$, a separation of $\Delta^2$ corresponds to a set of points that lies $2 N \Delta^2 /3 -\Delta^2 \approx  2 N \Delta^2 /3$ or $\sqrt{5N/7}\approx\sqrt{N}$ standard deviations from the mean.  In fact, this far out into the tail we cannot trust the Central Limit Theorem result: the tails of $P(r^2)$ are suppressed relative to the Gaussian form, since $0<r^2 < N \Delta^2$ while $P(r^2)$  computed from (\ref{PRCLT}) is non-zero outside these regions.

Instead, we can estimate $P(r^2)$ geometrically when $r^2$ is far below the mean. A hypersphere of radius $\delta$ surrounding a point has volume 
\begin{equation}
S_N = \frac{2  \pi^{N/2} }{N \Gamma\left(\frac{N}{2}\right) } \delta^N \, , 
\end{equation}
where we are allowing the volume of interest to have a radius $\delta$ which differs from the size of the box that contains the landscape.  Conversely, the landscape itself is (by assumption) a 
cube with volume
\begin{equation}
C_N = (2 \Delta)^N \, .
\end{equation}
Taking the ratio of these two quantities, the odds of randomly choosing two points in the landscape and finding that they are separated by a distance $r^2 < \Delta^2$ is $S_N/C_N$. This quantity rapidly approaches zero as $N$ becomes large for fixed $\Delta$, thanks to the familiar result that the N-cube is vastly larger than the N-sphere.  Quantitatively
\begin{equation} \label{pdelta}
P(r^2 <\delta^2) = \frac{S_N}{C_N} \sim \frac{1}{\sqrt{\pi N} }
\left[ \frac{e }{2 N} \left(\frac{\delta}{\Delta}\right)^2 \right]^{N/2} \, .
\end{equation}
Moreover, this is actually an {\em overestimate\/}, since it assumes the whole $N$-sphere fits inside the $N$-cube.

Recalling that we have $(\alpha/2)^N$ minima, we can therefore form $(\alpha/2)^{2N}/2$ distinct pairs of minima.  In this case, the expected number of pairs of minima whose separations are less than the characteristic scale of the landscape, $\Delta$ is 
\begin{equation}
\N(r^2<\Delta^2) =  \frac{1}{\sqrt{ 2 \pi N}} \left[ \frac{e }{2 N} \left( \frac{\alpha}{2}\right)^4   \right]^{N/2} \, .
\end{equation}
Thus we can quickly deduce:
\begin{equation}\label{flatineq}
\alpha \lsim 2N^{1/4} \Rightarrow \N(r^2<\Delta^2)  \ll 1 \, .
\end{equation}
If the inequality in (\ref{flatineq})
is satisfied, all the vacua in the landscape will be separated by a distance greater than $\Delta$. If we had looked for genuinely close points (since $\Delta$ is a substantial distance within the the landscape) the constraint would have been correspondingly tighter.   Numerically, if $N \sim 100$, $\alpha \gsim 6$ will give  a dense set of minima, whereas with $N\sim 500$ would need $\alpha \gsim 9$.   Consequently, it seems likely that we can regard all the extrema as being well separated from one another, and  thus forming a very sparse set within the landscape.   

We can also ask about the likely separation between inflationary saddles. In general, there $\alpha^N$ saddles in the landscape, but if we want  to concatenate two periods of inflation without a significant interruption, we want to end one trajectory and find that another saddle is within a distance $H$ inside field space. Since we have argued that the energy scale of the landscape is necessarily reduced by a factor of $1/\sqrt{N}$ relative to the Planck scale, $H \sim \delta \sim M^2/\mpl$, and $\delta /\Delta  \lesssim M^2/\mpl^2 \sim 1/N$.   Inserting these values into equation~(\ref{pdelta})
\begin{equation}
P(r^2 <H^2) =  \frac{1}{\sqrt{\pi N} }
\left[ \frac{e }{2 N} \frac{1}{N^2} \right]^{N/2} \, .
\end{equation}
 and the likelihood of connecting two inflationary saddles {\em by chance\/} drops super-exponentially as $N$ increases.   Even though the number of extrema grows exponentially with $N$, the ratio volumes between the hypercube and hypersphere grows super-exponentially with $N$. Moreover, the separation of scales needed to protect the minima grows with $N$, and since this is inversely proportional to the scale of inflation, the likelihood of finding two adjacent inflationary trajectories is further suppressed. 
This conclusion applies to both ``chained inflation'' \cite{Freese:2004vs} and ``folded inflation'' \cite{Easther:2004ir}.  Since we are  looking at a purely random potential, it is possible that both of these models might be realized in the context of the actual string landscape. On the other hand, multistage models of the form discussed in \cite{Burgess:2005sb} are safe, since the inflationary stages are well separated.     However, they are not rendered natural {\em simply\/} by the large number of extrema that the landscape contains, but must be justified by showing that in the actual string landscape there are well-defined subspaces where the density of extrema is much higher than in the rest of the volume. This problem is particularly acute for chained inflation, as the successive tunneling events it needs will be strongly suppressed if the minima are well separated as the barrier between them is then likely to be high.

\section{Conclusions}

The purpose of this paper has been to explore what follows {\em solely\/} from the large dimensionality of the string landscape, rather than from its detailed mathematical properties. 

 As was stressed earlier, the results here apply to general ``random'' potentials, but it is entirely possible that the actual  distribution of minima in the string landscape is  sufficiently non-random to thwart the arguments made in this paper.  However, these results will let us  distinguish between intrinsically stringy effects, and phenomena expected to arise in any physical model with a large number of scalar fields.  Even at this point, we can identify a number of ways in which string theory may yield a distribution of vacua which differs from those posited by our analysis.    For instance,  \cite{Giddings:2001yu} examines no-scale models where the potential is non-negative and the Hessian matrices  positive semi-definite. There are flat directions at tree-level but no tachyons, leading to a set of Hessian matrices that differs sharply from those considered here.   After SUSY breaking, the extrema seen in these models \cite{Denef:2004ze} still have Hessian matrices whose detailed structure ensures that the fraction of extrema which are actually minima exceeds  $1/2^N$.  Finally, we can seek out sub-regions within the landscape which contain an extra scale $M_\star$, so the separation of vacua in field space is controlled by $M_\star$ rather than $\mpl$.  Conversely, we might well find that variants of the arguments developed here still apply {\em within\/} the subspaces of the landscape in which the vacua are ``over dense'' or otherwise atypical.   
 
Interestingly,   while the number of minima produced by the simplest model -- a large set of non-interacting fields -- can be estimated trivially, turning on cross-coupings quickly changes the picture. In particular, if the cross terms between the fields have the same size as the diagonal quadratic terms, we can use random matrix theory to show that almost all extrema of the landscape are saddle points, and that there will be few or even no minima.  In the limit where the cross-terms between the fields are naturally suppressed relative to the self-interaction terms, we can quantify the amount of suppression  required  to ensure that the minima are not removed by these interactions.    

Finally, the distances between minima in the landscape are typically substantial, even though the {\em number\/} of minima grows exponentially with the dimensionality of the landscape.  This has immediate consequences for models such as chained inflation \cite{Freese:2004vs} or folded inflation  \cite{Easther:2004ir}, which concatenate several separate periods of inflation.   These models are not ruled out, but the necessary combinatorics cannot be provided simply by an appeal to the dimensionality of the landscape.  

We see several possible extensions of the approaches outlined in this paper, and feel that the appearance of random matrix theory in this context is particularly suggestive.  In particular, one of the authors [RE] and Liam McAllister are currently using  random matrix techniques to analyse the mass-spectrum found in models of N-flation \cite{Dimopoulos:2005ac}. 	The extent to which the properties of the actual string landscape matches the expectations derived solely from the large-N considerations explored here remains an open question. However, we have shown the dimensionality of the landscape alone is sufficient to draw conclusions about the likely set of extrema that exist within it.

 \section*{Acknowledgments}
RE is supported in part by the United States Department of Energy, grants DE-FG02-92ER-40704.  We thank Alan Guth,  Will Kinney,  Amanda Peet and Max Tegmark for useful conversations on this topic.  In particular, we are grateful to Douglas Stone for guiding us through the literature on random matrices, and to Shamit Kachru and Liam McAllister for detailed comments on a draft of this paper.  RE thanks the Aspen Center for Physics for its hospitality while some of this work was conducted.


\begin{thebibliography}{99}
 
\bibitem{Bousso:2000xa}
R.~Bousso and J.~Polchinski,
JHEP {\bf 0006}, 006 (2000)
[arXiv:hep-th/0004134].

\bibitem{Kachru:2003aw}
S.~Kachru, R.~Kallosh, A.~Linde and S.~P.~Trivedi,
Phys.\ Rev.\ D {\bf 68}, 046005 (2003)
[arXiv:hep-th/0301240].
 
\bibitem{Susskind:2003kw}
L.~Susskind,  ``The anthropic landscape of string theory,''
arXiv:hep-th/0302219.

 \bibitem{Susskindtalk}
 L. Susskind, 2003 ``Thoughts About Populating the Landscape''  Lecture Kavli Institute for Theoretical Physics, \url{ http://online.kitp.ucsb.edu/online/strings_c03/susskind/}

\bibitem{Douglas:2003um}
  M.~R.~Douglas,
  JHEP {\bf 0305}, 046 (2003)
  [arXiv:hep-th/0303194].
  
\bibitem{Ashok:2003gk}
  S.~Ashok and M.~R.~Douglas,
  JHEP {\bf 0401}, 060 (2004)
  [arXiv:hep-th/0307049].

\bibitem{Denef:2004ze}
  F.~Denef and M.~R.~Douglas,
  JHEP {\bf 0405}, 072 (2004)
  [arXiv:hep-th/0404116].
  
\bibitem{Giryavets:2004zr}
  A.~Giryavets, S.~Kachru and P.~K.~Tripathy,
  JHEP {\bf 0408}, 002 (2004)
  [arXiv:hep-th/0404243].

\bibitem{Liddle:1998jc}
  A.~R.~Liddle, A.~Mazumdar and F.~E.~Schunck,
  Phys.\ Rev.\ D {\bf 58}, 061301 (1998)
  [arXiv:astro-ph/9804177].

\bibitem{Arkani-Hamed:2005yv}
N.~Arkani-Hamed, S.~Dimopoulos and S.~Kachru,
arXiv:hep-th/0501082.

\bibitem{Distler:2005hi}
J.~Distler and U.~Varadarajan,
arXiv:hep-th/0507090.

\bibitem{Kobakhidze:2004gm}
  A.~Kobakhidze and L.~Mersini-Houghton,
  arXiv:hep-th/0410213.

\bibitem{Mersini-Houghton:2005im}
  L.~Mersini-Houghton,
  Class.\ Quant.\ Grav.\  {\bf 22}, 3481 (2005)
  [arXiv:hep-th/0504026].

\bibitem{Holman:2005ei}
  R.~Holman and L.~Mersini-Houghton,
  arXiv:hep-th/0511102.

\bibitem{Banks:1995dp}
T.~Banks, M.~Berkooz, S.~H.~Shenker, G.~W.~Moore and P.~J.~Steinhardt,
Phys.\ Rev.\ D {\bf 52}, 3548 (1995)
[arXiv:hep-th/9503114].

\bibitem{Binetruy:1996xj}
P.~Binetruy and G.~R.~Dvali,
Phys.\ Lett.\ B {\bf 388}, 241 (1996)
[arXiv:hep-ph/9606342].

\bibitem{Dvali:1998pa}
G.~R.~Dvali and S.~H.~H.~Tye,
Phys.\ Lett.\ B {\bf 450}, 72 (1999)
[arXiv:hep-ph/9812483].

\bibitem{Burgess:2001fx}
C.~P.~Burgess, M.~Majumdar, D.~Nolte, F.~Quevedo, G.~Rajesh and R.~J.~Zhang,
JHEP {\bf 0107}, 047 (2001)
[arXiv:hep-th/0105204].

\bibitem{Burgess:2001vr}
C.~P.~Burgess, P.~Martineau, F.~Quevedo, G.~Rajesh and R.~J.~Zhang,
JHEP {\bf 0203}, 052 (2002)
[arXiv:hep-th/0111025].

\bibitem{Jones:2002cv}
N.~Jones, H.~Stoica and S.~H.~H.~Tye,
JHEP {\bf 0207}, 051 (2002)
[arXiv:hep-th/0203163].

\bibitem{Arkani-Hamed:2003mz}
N.~Arkani-Hamed, H.~C.~Cheng, P.~Creminelli and L.~Randall,
JCAP {\bf 0307}, 003 (2003)
[arXiv:hep-th/0302034].

\bibitem{Kachru:2003sx}
S.~Kachru, R.~Kallosh, A.~Linde, J.~Maldacena, L.~McAllister and S.~P.~Trivedi,
JCAP {\bf 0310}, 013 (2003)
[arXiv:hep-th/0308055].

\bibitem{Silverstein:2003hf}
E.~Silverstein and D.~Tong,
Phys.\ Rev.\ D {\bf 70}, 103505 (2004)
[arXiv:hep-th/0310221].

\bibitem{Alishahiha:2004eh}
M.~Alishahiha, E.~Silverstein and D.~Tong,
Phys.\ Rev.\ D {\bf 70}, 123505 (2004)
[arXiv:hep-th/0404084].

\bibitem{Hsu:2003cy}
J.~P.~Hsu, R.~Kallosh and S.~Prokushkin,
JCAP {\bf 0312}, 009 (2003)
[arXiv:hep-th/0311077].

\bibitem{Firouzjahi:2003zy}
H.~Firouzjahi and S.~H.~H.~Tye,
Phys.\ Lett.\ B {\bf 584}, 147 (2004)
[arXiv:hep-th/0312020].

\bibitem{Hsu:2004hi}
J.~P.~Hsu and R.~Kallosh,
%
JHEP {\bf 0404}, 042 (2004)
[arXiv:hep-th/0402047].

\bibitem{Iizuka:2004ct}
N.~Iizuka and S.~P.~Trivedi,
Phys.\ Rev.\ D {\bf 70}, 043519 (2004)
[arXiv:hep-th/0403203].

\bibitem{DeWolfe:2004qx}
O.~DeWolfe, S.~Kachru and H.~L.~Verlinde,
JHEP {\bf 0405}, 017 (2004)
[arXiv:hep-th/0403123].

\bibitem{Dvali:2003vv}
G.~Dvali and S.~Kachru,
arXiv:hep-th/0309095.

\bibitem{Easther:2004qs}
R.~Easther, J.~Khoury and K.~Schalm,
JCAP {\bf 0406}, 006 (2004)
[arXiv:hep-th/0402218].

\bibitem{Blanco-Pillado:2004ns}
J.~J.~Blanco-Pillado {\it et al.},
JHEP {\bf 0411}, 063 (2004)
[arXiv:hep-th/0406230].

\bibitem{Easther:2004ir}
R.~Easther,
%
arXiv:hep-th/0407042.

\bibitem{Freese:2004vs}
K.~Freese and D.~Spolyar,
JCAP {\bf 0507}, 007 (2005)
[arXiv:hep-ph/0412145].

\bibitem{Burgess:2005sb}
C.~P.~Burgess, R.~Easther, A.~Mazumdar, D.~F.~Mota and T.~Multamaki,
JHEP {\bf 0505}, 067 (2005)
[arXiv:hep-th/0501125].
 
\bibitem{McAllister:2005mq}
L.~McAllister,
%
arXiv:hep-th/0502001.
\bibitem{Cremades:2005ir}
D.~Cremades, F.~Quevedo and A.~Sinha,
 ``Warped tachyonic inflation in type IIB flux compactifications and the
%
arXiv:hep-th/0505252.

\bibitem{Dimopoulos:2005ac}
  S.~Dimopoulos, S.~Kachru, J.~McGreevy and J.~G.~Wacker,
  arXiv:hep-th/0507205.

\bibitem{Easther:2002rw}
  R.~Easther and W.~H.~Kinney,
  Phys.\ Rev.\ D {\bf 67}, 043511 (2003)
  [arXiv:astro-ph/0210345].
  
\bibitem{Tegmark:2004qd}
  M.~Tegmark,
  JCAP {\bf 0504}, 001 (2005)
  [arXiv:astro-ph/0410281].

\bibitem{Easther:2005nh}
  R.~Easther and J.~T.~Giblin,
  arXiv:astro-ph/0505033.

\bibitem{Fyo}
  Y.~V.~Fyodorov, Phys. Rev. Lett. {\bf 92}, 240601 (2004);  Erratum {\bf 93} 149901(E) (2004)

\bibitem{Linde:2004kg}
A.~Linde,
Phys.\ Scripta {\bf T117}, 40 (2005)
[arXiv:hep-th/0402051].
\bibitem{Cline:2005zy}
J.~M.~Cline,
arXiv:hep-th/0501179.

\bibitem{Mehta}  M. L. Mehta, 1990 {\em Random Matrices\/}, 2nd Edition, Academic Press, Boston

\bibitem{Tracy:1995xi}
  C.~A.~Tracy and H.~Widom,
  Commun.\ Math.\ Phys.\  {\bf 177}, 727 (1996)
  [arXiv:solv-int/9509007].
  
 \bibitem{Schott} J R Schott, 2005 {\em Matrix Analysis for Statistics\/}, Wiley.
 
\bibitem{Copeland:1994vg}
E.~J.~Copeland, A.~R.~Liddle, D.~H.~Lyth, E.~D.~Stewart and D.~Wands,
Phys.\ Rev.\ D {\bf 49}, 6410 (1994)
[arXiv:astro-ph/9401011].
 
\bibitem{Kallenberg} O.~Kallenberg, {\em Foundations of Modern Probability\/}, Springer 2002.

\bibitem{forth1} R.~Easther, W.~Kinney and B.~Powell.  In Preparation.
 
\bibitem{Giddings:2001yu}
S.~B.~Giddings, S.~Kachru and J.~Polchinski,
Phys.\ Rev.\ D {\bf 66}, 106006 (2002)
[arXiv:hep-th/0105097].
 
\end{thebibliography}
 \end{document}